# Analysis of the Smart Grid as a System of Systems

Tawfiq Aljohani
Energy Systems Research Laboratory, Florida International University, Miami, FL
Taljo005@fiu.edu

## Abstract

The energy grid is currently undergoing a historic change of state from the traditional structure where a utility owns the generation, transmission and distribution services into an integrated smart grid in a monopolistic market which introduce consumers as active players in managing and controlling the power. This evolution adds more complexity to the energy scene as it requires an unprecedented partnership of different fields and technologies to establish successful integration of components with a bidirectional transfer of information and energy. The Smart Grid Interoperability Panel (SGIP) was incorporated in late 2009 to oversee the efforts in defining a set of standards and interoperability layers to ensure a secure integration and operation of various elements of the system. This paper discusses the smart grid concept as a system-of-systems (SoS); analyzes the interoperability framework models developed by SGIP and the European Union Smart Grid Cooperation Group respectively; presents the stakeholders as domains and zones associated with the interoperability layers; and illustrates other architectural concepts such as fault-tolerance, software integration platform and design characteristics of smart grid as an elegant system.

## Abbreviations

**SoS:** System of Systems

**SGIP:** Smart Grid Interoperability Panel.

**AMI**: Advanced Metering Infrastructure.

**CIM**: Computer Integrated Manufacturing.

**EPRI**: Electric Power Research Institute.

**NIST**: National Inst.of Standard and Tech.

**GWAC**: GridWise Architecture Council.

**CAISO**: California Independent System Operator.

**DR:** Distributed Resources.

**FTP:** File Transfer Protocol.

**ICT**: Information & Communication Tech.

**DSM**: Demand-Side Management.

**SOA**: Service-Oriented Architecture.

**IoT**: Internet of Things.

**EISA**:Energy & Independence Security Act.

## A. Introduction

The electrical power grid is considered as the most complex system humanity ever built. It has enabled improved living standards for humans where all life aspects are directly linked with energy consumption, adding more complexity to the power industry over the time. The continuous surge in power consumption is proportional to several stochastic factors such as the increase in population, needs for luxurious lifestyle as well as the ongoing technological advancement that drive human behavior toward uncertainty. This, of course, comes with a cost of devastating our environment. Therefore, the concept of the smart grid has emerged in the past decade to call for proper integration and utilization of resources with the inclusion of consumers as active players in the production, control, and decision-making of power.

The transition from the traditional vertically integrated hierarchy of the energy companies into this form of "smart, intelligent grid" or "energy grid with brain" is no doubt a game change that will not be possible without the effective participation of the information and communication technologies to enable an instantaneous, all time, safe two-way passage of information among the linked parties. It requires a systematic integration of the most advanced capabilities in the communication, data sensing and measurement and information technology to successfully transform the current scene from hundreds of centralized power sources, which supply receiving-end consumers with energy flowing in a one way direction only, into what potentially could be tens of millions of plug-in electric cars, thousands of distributed sources, demand response control and management programs, and other consumer-owned assets that constitute the concept of a smart grid. Figure 1 shows an Electric Power Research Institute (EPRI) description of such transition [1].

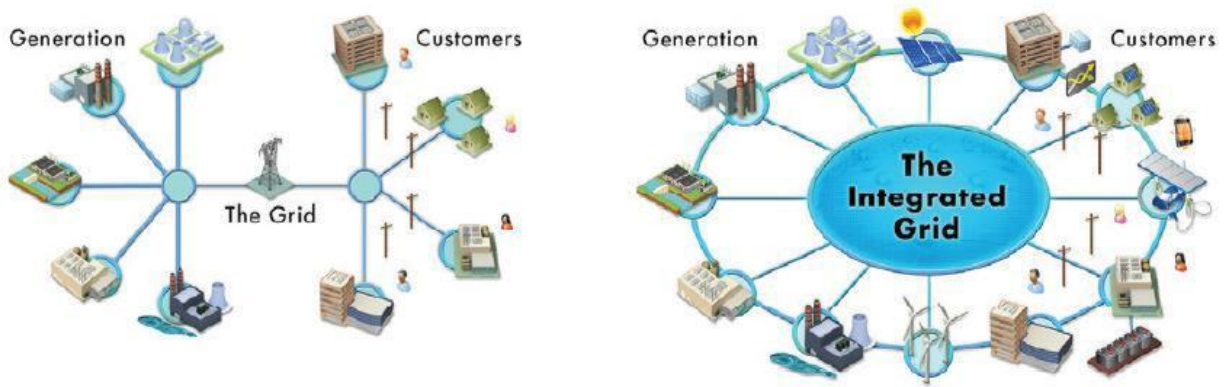

Figure 1: The proposed transition from a traditional into an integrated "smart grid"

**B. The Smart grid as a System of Systems (SOS)**

The smart grid can be considered as a SoS based on the attributes that define what constitute SoS. According to [2, 3], the characteristics of SoS are as follow:

- **Operational Independence of the Elements**: A single component or sub-system of a smart grid can operate effectively even if isolated from the overall network.
- **Managerial Independence of the Elements**: All components in the smart grid operate independently, although they together influence the result of the whole network.
- **Evolutionary Development**: Elements of the smart grid can be isolated, added, integrated or retired at any time without causing an impact to other parts of the electrical system. The essence of the smart grid is its ability to perform this feature specifically for an efficient utilization and control of resources.
- **Emergent Behavior**: The integration of different components with various capabilities to serve stochastic energy loads leads to emergent properties that fulfill the major purpose of the smart grid, which is mainly to meet the demand in a clean, economical, and efficient way with the participation of multiplayer.
- **Geographic Distribution**: The elements of the smart grid are geographically dispersed. They are linked only through information exchange channels.

There is a need to adopt a system architectural thinking to allow for an optimal synergy between components of the smart grid for secure data exchange and management. The scattering of its elements geographically makes it challenging to collect reliable data, which are collected using intelligent devices that require secure infrastructure to make the communication channels well-prepared not only to gather data from decentralized systems but also to filter, analyze, and coordinate them in a centralized data warehouse.

Therefore, there is no better option for the smart grid's architect but to utilize the SOS concept to develop a systematic set of requirements and standards and to establish a common liaison among stakeholders to accelerate the erection of the smart grid. The essence of the to-be-established standards is to enable an interoperable operation to support proper interfacing among the various elements of the system.

**B.1 The Interoperability of the Smart Grid**

The Smart Grid Interoperability Panel (SGIP) was established by the National Institute of Standards and Technology (NIST) in late 2009, in response to the Energy Independence and Security ACT (EISA) of 2007 which charged NIST with the "***Primary responsibility to coordinate development of a framework that includes protocols and model standards*** *for information management to achieve interoperability of Smart Grid devices and systems*" [4].

Among the missions of SGIP [5]:

- Oversee the concerns of the associated stakeholders,
- Force quick fulfillment of smart grid goals,
- Intermediate of "Internet of things (IOT)" in energy arena,
- Set up standards for efficient integration of different resources,
- Ensure future energy sustainability.

The smart grid is a complex system that is to-be-built over existing electrical infrastructures which require different layers of interoperability. NIST has considered the eight layers model developed by the GridWise Architecture Council (GWAC) [6], shown in figure 2.

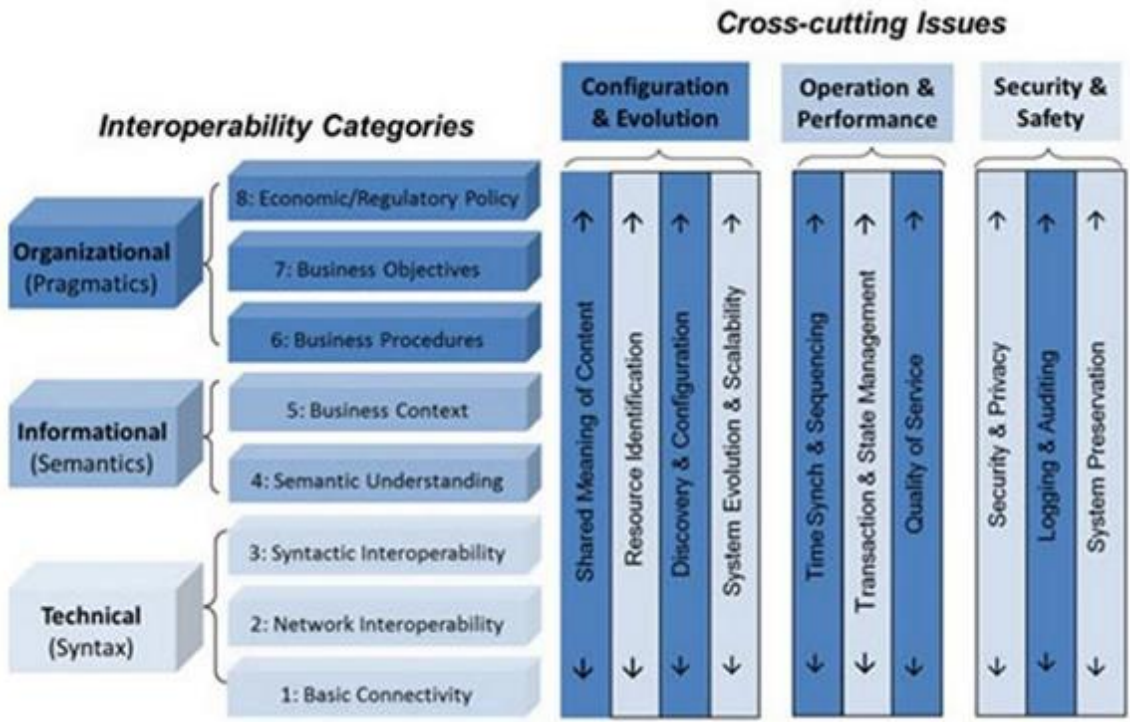

Fig 2: GWAC defines interoperability and information exchange requirements in eight layers [7].

The main goal of the interoperability, in the context of the smart grid, is to facilitate secure transition and interaction of various systems to ensure the reliability, sustainability, efficiency, economy and security of operation. Sievers and Madni [8] define the interoperability in general as the method of which two systems or more can interface and exchange information efficiently to achieve an overall goal.

GWAC's aim of developing the framework, shown in figure 2, is to initiate a background to "identify and debate" the issues related to interoperability which will eventually simplify the integration by defining the cross-cutting issues needed to be addressed. As seen in figure 2, GWAC categorize the interoperability of smart grid into:

- **Technical (syntax) Interoperability:** Which is required to establish physical and logical connections among elements, to exchange messages through a variety of networks, and to provide syntactic interoperability for a common understanding of data structure in the shared information.
- **Informational (Semantics) Interoperability:** To ensure semantic understandings of the context of the exchanged information among the various integrated components, and to improve business knowledge based on these analyzed data.
- **Organizational (Pragmatics) Interoperability:** To calibrate operational business process and procedures, to plan strategic and tactical common objectives among stakeholders, and to ensure understanding and abidance of economic and political policies and requirements.

The cross-cutting issues provided in the framework can mitigate interdependencies within the electrical structure as well as with other elements of the smart grid [7]. It is tailored to present documents for audiences of different interest, which offer flexibility in aligning the stakeholders based on their mutual targets, leading to enhance dialogue and improve opportunities as well as ability to absorb technical innovations introduced in the future.

However, GWAC model, sponsored by the Department of Energy, is not the only model found to frame the interoperability in this arena. Many models have been proposed and funded by governmental or private research grants. Figure 3 shows a framework developed by the European Union Smart Grid Coordination Group that distinguishes the various layers of interoperability discretely [9]. This framework defines the integration of the smart grid as a system-of-systems based on five layers of applications within three dimensions: the domains (described by the electrical infrastructure), the information zones, and the interoperability dimensions. These five layers are business, function, information, communication and component, respectively. Each layer deals with a different perspective of the smart grid, spanned by an electrical domain and information zone, to allow clear representation of entities and their relationships by the interoperability facet. Discussion on this model is presented in section 4.1 of this work.

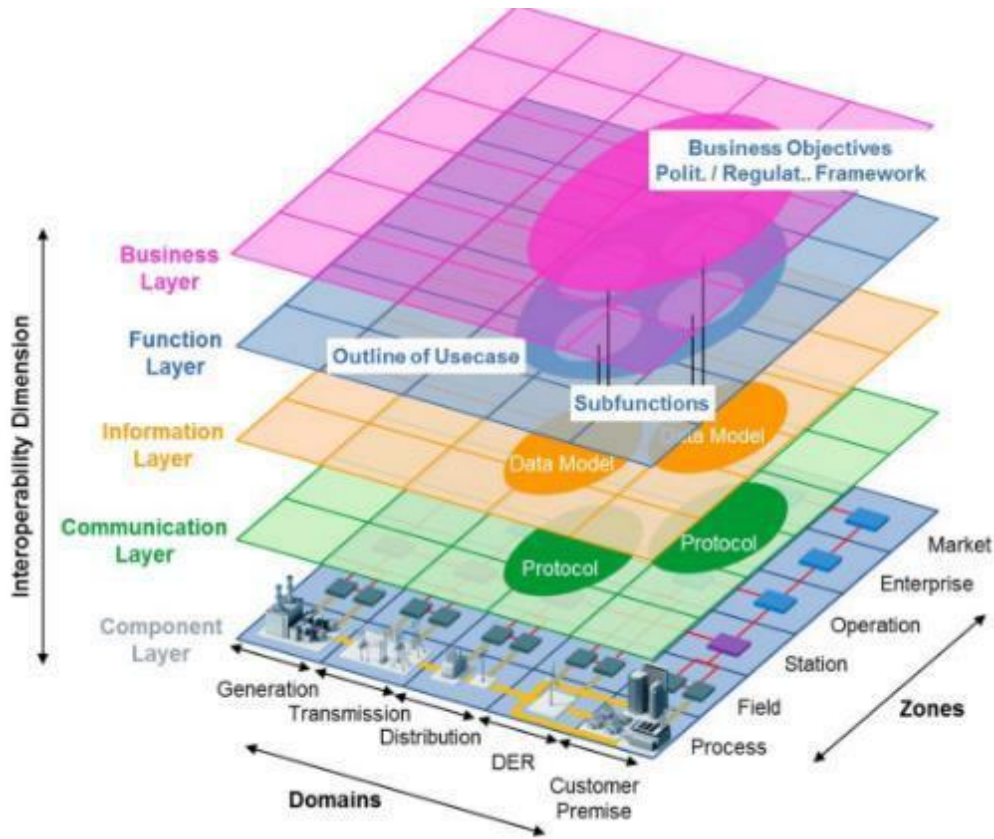

Figure 3: European Smart Grid Model dissects interoperability into five discrete layers [9]

**B.2 Structure, Integration and Ontology of the Smart Grid**

Sievers and Madni [3] give an explanation of the terminology "ontology" in the view of the concept of a SoS as a defining set of objects and respective relationships to better analyze system's domain knowledge, eliminate confusions and promote semantic integration. This concept is what is needed for processing the massive amount of real-time data collected from involved elements of the smart grid to achieve a successful integration.

The domains in this study are physical infrastructures correspondent to the power grid, arranged in a hierarchal aspect of bidirectional service providers as shown in the bottom of figure 3, from the generation at the top of the chain to the customer premise, who act as a part of demand-side management programs. This means that the service provider can be located at any segment in the smart grid plane, and has particular tasks to perform facilitated by the interoperability model. We highlight in this paper the associated stakeholders of any smart grid from the view of the domains and zones only. The EU Smart Grid Group has utilized NIST's definition of the smart grid domains in its interoperability model [10]:

- **Bulk Generation**: responsible for producing energy in bulk quantities. It includes the traditional power stations run by fossil fuel, nuclear, renewables …etc.
- **Transmission**: the infrastructure for transmitting the energy over long distances.
- **Distribution**: considered as the domain of which smart grid is massively transforming its shape. It is responsible for delivery of energy to consumers and is expected to witness significant modification in infrastructure to allow for customer participation in load management and the integration of dispersed resources.
- **Distributed Resources (DR)**: the DRs allow the implementation of features sought by the smart grid. It has technical and dynamical limitations that need to be taken into consideration for successful integration with the distribution network. It is a subject still undergoing intense study at policy and requirement levels [11].
- **Customer Premises**: major players in the concept of the smart grid, actively participating as load consumers, producers and managers. It covers wide variety of classes such as residential, municipal, industrial, commercial …etc.

On the other hand, the zones, depicted as cross-cutting issues in figure 2 and as a dimension in figure 3, illustrate the hierarchy of power system management in the smart grid model. Reference [9] utilizes the Purdue Reference Model for Computer Integrated Manufacturing (CIM) [12] to build the foundation of energy management in the smart grid as enterprise architecture in multiple layers and stages.

The ultimate goal of "partitioning the zones" is to allow different functions to be employed at respective zones. For example, real-life functions can be dictated, as seen in the zones dimension of figure 3, in both the field and station zones to perform given duties such as smart-metering, automation, protection and so on. Similarly, functions that require the interaction of the multiple players (i.e. generation and maintenance scheduling, weather forecasting, substations coordination) can be located using the operation zone. The following zones are described as following:

- **Process**: the process of energy conversion using various technologies, along with linked apparatus (i.e. transformers, lines and conductors, sensing devices, loads).
- **Field**: include the switchgear and protection for power system apparatus.
- **Station**: organize the control and management of power in the respective domains (i.e, plug-in electric cars and demand side management in the DER domain, energy management system in the generation and transmission domains).
- **Enterprise**: include asset management, representation of interests for energy providers, utilities, traders... etc.

- **Market**: retail and mass market, analysis of production vs. consumption ..., etc.

As mentioned in section 2 of this report, to obtain desired functionality of the concept of the smart grid is entirely based on advances in the information and communication technologies (ICTs). Reference [13] cites a customer-based integration platform shown in figure 4. The conceptual design of this platform is to collect data from consumers through smart meters, famously known as Advanced Metering Infrastructure (AMI), and centralize them into a secure data warehouse for further management and analysis. The AMI enables channeling precise information which gives the service provider the advantage of utilizing dynamic pricing rates based on both the type and time of consumers' consumption.

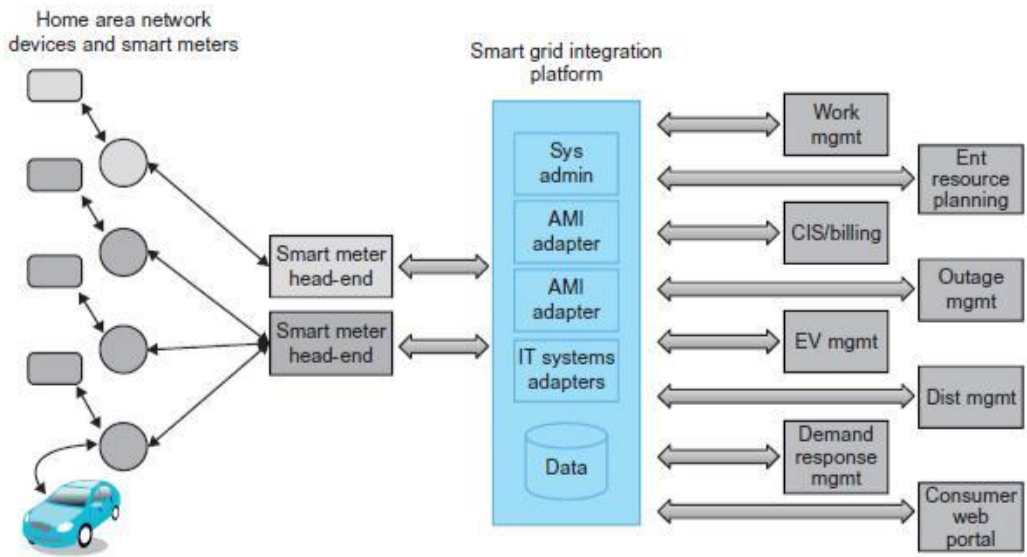

Figure 4: Communication network and data channeling using smart meters.

The concept of the smart grid as a SoS need to loosely connect all elements while tolerating future modifications without impacting the linked systems. This kind of architecture is achieved through Service-Oriented Architecture (SOA), illustrated in details in reference [14]. For further explanation of the need for proper interoperability at the software integration level, refer to figures 2 and 3 of California Independent System Operator (CAISO) 2020's interface architecture, based on utilizing the CIM to analyze the interfacing across domains mentioned in 3.1 of this paper, in addition to HAN devices, AMIs, and other communication technologies which are integrated through secure file transfer protocol (FTP), managed via web services [15].

## B.3 SOA as Middleware in the Smart Grid

Service Oriented Architecture (SOA) is a software model with the objective of providing efficient, reusable and flexible heterogeneous services to the end users. It has the potential of providing interoperable management of components via extensible markup language (XML) protocol, where all data provided from each element in the smart grid is converted then exchanged in this format. In addition, SOA allows utilization of the existing (aka legacy) electrical systems in the overall service of the smart grid, as well as the capability of integrating new and heterogeneous components and services over the time in an enterprise environment, ensuring reliable and securable links of data resources, business process, platforms and applications.

SOA as a middleware focuses on the stacks of protocol and schemes for scheduling the services efficiently to fulfill the requests of various users. It is intended to form a "bridge" that links the operator of the requested services with its associated clients and systems. Once information of a particular service is received, the middleware allows the operator to dynamically supply the requested service based on its nature and characteristics via the AMIs [16]. Specifically, when a user accesses the grid for a particular service, one of the main missions of SOA is to locate and communicate with the relevant service providers. In case the demanded service is not provided by any service provider, then SOA highlights the need for the integration of this service into the overall system [17]. Such adoption of middleware structure makes it easier to incorporate multiple open interfaces and interconnection of heterogeneous networks and devices.

One of the key concepts of the smart grid is to allow energy consumption based on real-time rates. The generation and transmission of such data are central to the successful implementation of SOA to serve the power network. Therefore, the SOA needs to have the capability to dynamically offer data interpretation of end-user equipment, size the time and amount of flow of information from users, and able to flexibly exchange information with multiple types of smart meters. To achieve that successfully, Zho suggests decoupling of the middleware into several applications based on their functionalities, as seen in figure 5, which eventually allows dynamical interaction between energy companies and consumers with regard to type of services requested and provided, allowing AMIs, through wide area networks (WANs), to optimize energy consumption via controlling the smart appliances based on previous agreements between related parties.

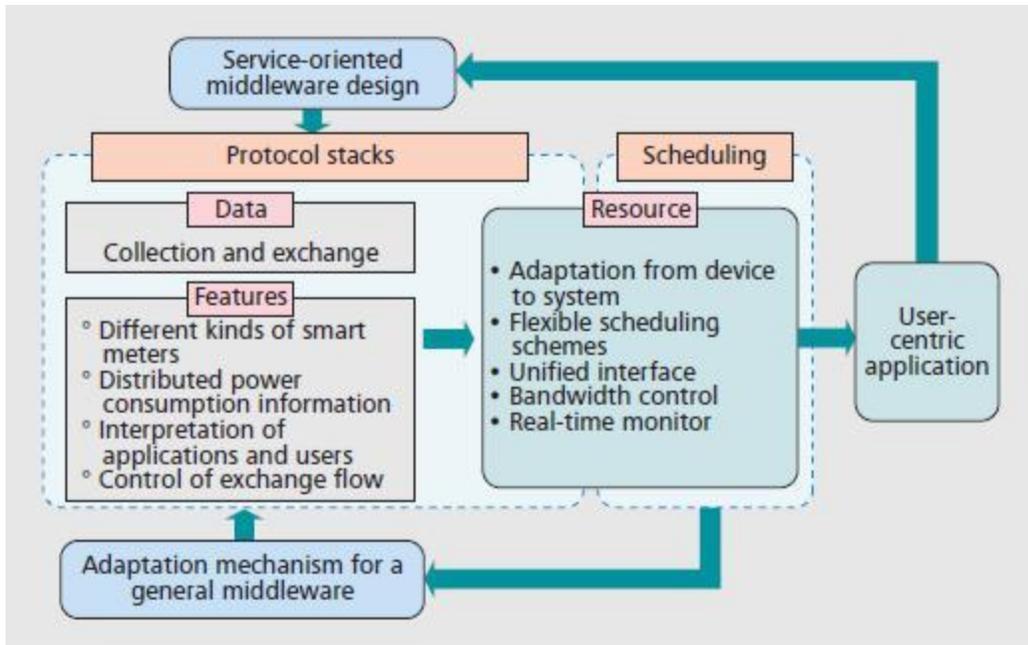

Figure 5: Middleware design as multiple application functionalities [16].

Reference [16] adopted three main layers, service-driven middleware to provide heterogeneous services in the context of the smart grid, as seen in figure 6:

- **Transmission Part**: a part of the middleware that works as a data layer. It arranges the amount of energy needed from the generation points to the distribution (consumers-end) points via AMIs which send the clients' appliances consumption to a data warehouse that influences a controller (owned by the utility) which is designed to manage the shifting of power consumption in accordance with the price of available energy.
- **Control Part**: works as device management that connect the transmission part with the user part. It is considered as modules of exchange orders among the other two parts.
- **User Part**: provides reliability, security, the bandwidth of providing the services. Specifically, data about consumption arrives at the controller safely and discreetly. The middleware uses secure methods of communication that encode and encrypt information that is passed in the form of XML to ensure the privacy of consumers.

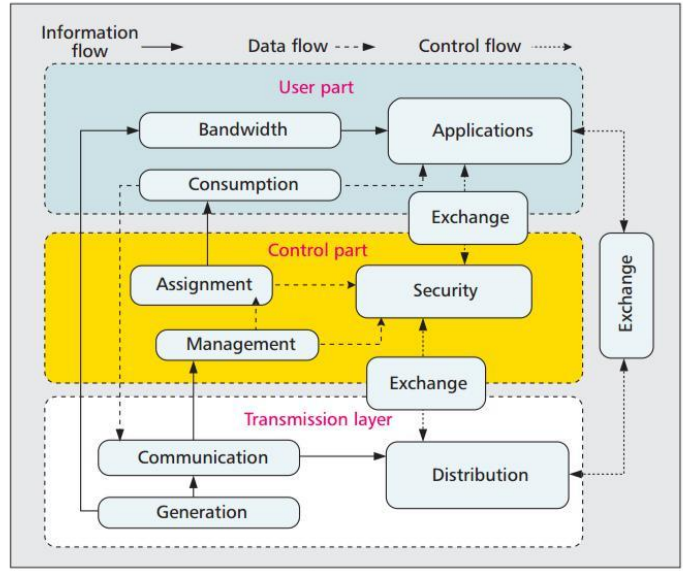

Figure 6: A designed middleware to provide heterogeneous services in the smart grid

However, we note that there are certain shortcomings in using this architecture in the smart grid platform. The first limitation is that it is not useful for standalone applications that have no components to be integrated with other components. Also, this SOA architecture will be useless if it does not support the message-passing mechanism needed to establish a successful connection. Moreover, such design of a middleware is also not suitable for the integration of services that have short time duration of their execution. Specifically, it will need ultra-advanced technologies to cope with the unpredictability of renewable energy sources, which varies in minute-to-minute and is mainly depended on weather forecasting to allocate energy needs in the power market better. Further, the SOA is not suitable for the integration of services which are developed in a homogeneous environment since they are using the same language, tools, and techniques; therefore such services can interact easily without the need for paying additional costs to incorporate each of them in the middleware.

A major drawback of using middleware is that massive data processes will result in exponential connections to the server which will overload it. Moreover, in this architecture, messages are transmitted and received in the XML format that enlarges its size; as a result, more bandwidth is consumed, and extra resources are utilized. Also, in the case of the occurrence of several failures in the network at the same time, then it will be extremely hard to assess the reliability of the individual point using the SOA [18]. However, we recognize that these shortcomings do not affect the stance of this SOA as a key element to ensure the success of erecting smart grids in an enterprise platform.

There exist areas of improvement for the proper implementation of SOA in the context of the smart grid including solving the challenges in the interoperability of the software components, the scalability of the grid due to increased number of its users, the discovery of the service providers and consumers, the interaction of mobile consumers which could be millions in the case of EVs owners participation with the power network, the integration of services as well as the close interaction with the physical, data and business layers, the modernization of the topology of energy distribution grids, and the lack of standardization of mechanisms and tests that are required before integration of any components in the overall system.

**B.4 Design Characteristic of the Smart Grid as an Elegant System**

The proper inclusion of consumers is one of the key factors needed to successfully architect the smart grid. So how can such complex system be elegant and encourage people to participate in the so-called "the energy internet" and "the electricity with brain"? It would be by offering them the access to easy-to-use smart appliances as well as introducing them efficiently to its basic concepts of operation, which is the part known as demand-side management (DSM), which is the direct influence of the consumers on the operation of the smart grid. Straiger et al. (2010) develop a model to predict the consumers' possible adoption to cooperate and share their data usage. The model focuses on the "perceived usefulness" and "perceived ease-of-use" as two main factors to develop behavioral understandings of the acceptance of using the new technology. The author of the work realized that the adoption would be achieved if the ease-of-use of the new smart devices is as easy as those used before by the consumers, which raise the intention to use them. It is very interesting that the model noticed no correlation between the "perceived usefulness" factor and the consumers' intention of use, which make a conclusion of the absence of knowledge, or in other words, the interest to learn how the smart appliances work. Thus, without the ease-of-use path, consumers will never know nor able to use the potentials of the smart appliances [19].

Comparing this model with Salado's metrics described in [20, 21], the primary objective of the smart appliances is to provide the same service as the traditional ones, which smart appliances not only do but also in a more efficient and reliable way. A good example of a new smart appliance is the electric washer and dryer machine: i) It does the job faster and is less quit (functionality and performance); ii) Has appealing look compared to the old machines; iii) Consumers are incentivized to acquire them at discounted prices due to federal, state and private subsidies in particular locations (availability); iv) Certified as energy efficient which consume less power to do a better job (efficiency); v) Adapted to the resident autonomously and assisted in their ways of living (adaptability). The combination of these metrics produces an elegant system-of-systems that can be incorporated and integrated into the overall smart grid, where consumers can control their own power to manage their consumption based on the real-time energy rates offered to them, and as can be concluded from the models shown in Figures 2 and 3 of this report.

However, what could be the best approach to offering easy-to-use and –control smart appliances? There have been many brainstormed ideas; one find its way to reality is to take advantage of smartphones and tablets to offer the consumers the ability to easy and distanced control of their smart appliances via easy-to-use applications that can be installed and managed by the consumers at anytime and anyplace with access to the internet. Such applications have been developed and sponsored by most of the manufacturing companies such as General Electric, Whirlpool and others, where only four simple steps will be required: access the internet, download the application, create an account with the service provider, and register the appliance.

In overall, considering the design features of elegant systems illustrated by Sievers and Madni [20], we find that the architecture of the smart grid can have the same characteristic of an elegant system as following:

- **Interoperability**: Its concept is based on interfacing of various systems.
- **Innovation**: Ability to adopt innovation and accommodate them, integrate new technologies in energy, communication and information arenas.
- **Affordability**: Develop markets to allow multivendor acquisition of interoperable network. To target reducing prices of energy and capital costs.
- **Flexibility**: Allow unprecedented bidirectional power flow, participation of consumers to manage their load and thus bills …etc.
- **Sustainability**: The penetration of renewable sources is one of its essential characteristics. Significant reduction of toxic gases is expected as well.
- **Upgradability**: Allow the modification and upgrading of the grid components at any time without operational interruption.

**B.5  Human-System Integration in the Context of the Smart Grid**

Human-System Integration (HSI) is an interdisciplinary technical and managerial process that aims to integrate the human factor with the systems [22]. According to the US General Accounting Office, the human components account for more than half of the lifecycle costs of the large complex systems and has evolved as one of the crucial concepts to analyze better, operate and manage in reliable aspect those complex systems that were built by the humanity [23].

HSI offers the opportunity to integrate the human factor in the sociotechnical systems, to emphasize the advantage of the human capabilities while making sure to circumvent their potential limitations. Specifically for the smart grid, the US department of energy has highlighted the enabling of active participation of humans as one of its critical success and definitions. Whether by participating in DSM programs, owning hybrid or electric plug-in cars, acting as an independent energy producer using DR, or allowing the power utilities to collect their consumption data, including the type of watts consumed during specific hours of the day, we can see that the smart grid is indeed being built around the human factor.

Although a considerable amount of consumers are still in doubt about involving in such interaction, sharing dynamically updated consumption of energy through centric database would effectively help in addressing the reliability of the service and creating new trends of services that consumers could benefit from it. The decentralization of control of such data would encourage more customers to participate through the information layer in the interoperability model discussed in section 2 of this report. Such human interface would eventually be of great help in rapid and informed decision making.

## C. Conclusion

Successful implementation of the smart grid is essentially based on achieving brilliant architecture and design that can both accommodate the bi-directional exchange of power and information, and ensure proper coordination and management of all resources. Applying system thinking approach as well as an understanding of the smart grid as a system-of-systems, which embodies emergent behavior due to the interaction of various interfaces, is the ultimate path to successfully fulfill the goals of creating an intelligent, self-aware grid.

In addition to the urgency of establishing standards and guidance for the technical integration of smart devices, resources, and automation into the system; it is quite essential to reliably integrate consumers as well into the active participation in the operation. This should be achieved not only by building the required infrastructure to allow it but also by ensuring sufficient knowledge and awareness on the consumers' side. It is also of great importance to isolate the complexity of the grid away from them as well as business managers and investors, who might be technically limited, which is done via the successful design of the smart grid as an elegant system.

In this paper, we analyzed the concepts of the system-of-systems architecture and characteristics needed in the development and design of the smart grid. We highlighted the smart grid concept as a system of systems, analyzed recognized interoperability models, presented the stakeholders as domains and zones associated with the interoperability layers, and illustrated other architectural concepts such as fault-tolerance, software integration platform and design characteristics of the smart grid as an elegant system. To sum up, we suggest that the comprehension of the smart grid as a system-of-systems is the only possible path to successfully architect the system that is to be the most sophisticated system humankind has ever dreamed of so far: an intelligent, self-realizing power grid.

**Biography**

**Tawfiq M. Aljohani** received his B.SC degree in electrical engineering in 2009 from King Abdulaziz Univeristy, Jeddah, Saudi Arabia. He received two Master of Science degrees from the University of Southern California, Los Angeles, CA in electric power engineering (2014), and green technologies (2017) and a grad certificate in System Architecting and Engineering (2017). He is currently doing his Ph.D. at the Energy Systems Research Laboratory at FIU with research interests that include renewable energy integration and smart grid applications, system architecture applications in power systems, and electric vehicle operation and control.